\documentclass[12pt]{article}
\usepackage{graphicx}
\usepackage{hyperref}

\newcommand\pubnumber{DAMTP-2011-2}
\newcommand\pubdate{January 14, 2011}

\def\cambridge{$^a$DAMTP, University of Cambridge,
Wilberforce Road, Cambridge CB3 0WA, UK}
\def\sm{$^b$Department of Physics, College of William \& Mary, Williamsburg, VA 23187-8795, USA}
\def\supa{$^c$SUPA, School of Physics and Astronomy, University of Edinburgh, Edinburgh EH9 3JZ, UK}
\def\support{\footnote{Work supported by the UK Science and Technology Facilities Council.}}
\def\Speaker{Speaker. Current address: Institute of High Energy Physics, Chinese Academy of Sciences, 
Beijing 100049, China.}

\def\Title#1{\begin{center} {\Large #1 } \end{center}}
\def\Author#1{\begin{center}{ \sc #1} \end{center}}
\def\Address#1{\begin{center}{ \it #1} \end{center}}

\newcommand\pubblock{\rightline{\begin{tabular}{l} \pubnumber\\
         \pubdate  \end{tabular}}}
\newenvironment{Abstract}{\begin{quotation}  }{\end{quotation}}
\newenvironment{Presented}{\begin{quotation} \begin{center} 
             PRESENTED AT\end{center}\bigskip 
      \begin{center}\begin{large}}{\end{large}\end{center} \end{quotation}}

\def\beq{\begin{equation}}
\def\eeq#1{\label{#1}\end{equation}}
\def\eeqn{\end{equation}}

\def\beqa{\begin{eqnarray}}
\def\eeqa#1{\label{#1}\end{eqnarray}}
\def\eeqan{\end{eqnarray}}

\let\bar=\overbar

\def\Dslash{\not{\hbox{\kern-4pt $D$}}}
\def\dslash{\not{\hbox{\kern-2pt $\del$}}}

\def\msb{{\bar{\ssstyle M \kern -1pt S}}}

\begin{document}
\begin{titlepage}
\pubblock

\vfill
\Title{A lattice calculation of $B\rightarrow K^{(*)}$ form factors\support}
\vfill
\Author{Zhaofeng Liu$^a$\footnote{\Speaker}, Stefan Meinel$^b$, Alistair Hart$^c$, Ron R. Horgan$^a$,\\
Eike H. M\"uller$^c$, Matthew Wingate$^a$}
\Address{\cambridge\\
\sm\\
\supa}
\vfill
\begin{Abstract}
Lattice QCD can contribute to the search for new physics in $b \to s$
decays by providing first-principle calculations of $B \to K^{(*)}$
form factors. Preliminary results are presented here which complement sum
rule determinations by being done at large $q^2$ and which improve upon
previous lattice calculations by working directly in the physical $b$
sector on unquenched gauge field configurations.
\end{Abstract}
\vfill
\begin{Presented}
The 6th International Workshop on the CKM Unitarity Triangle\\
University of Warwick, UK, 6-10 September 2010
\end{Presented}
\vfill
\end{titlepage}
\def\thefootnote{\fnsymbol{footnote}}
\setcounter{footnote}{0}

\section{Introduction}

The $b\rightarrow s$ flavour changing neutral current transition is suppressed in the standard model.
Dominant contributions to rare $B$ decays such as $B\rightarrow K^{(*)}l^+l^-$ come from loop diagrams: 
box and penguin diagrams. These rare $B$ decays are good windows for looking for new physics: New particles
beyond the standard model could appear in the loops and change the decay widths of those rare decays.

The starting point of theoretical calculations of weak decays of hadrons is the effective weak Hamiltonian.
In the standard model, there are ten operators in the effective Hamiltonian for radiative and semileptonic decays.
The dominant short distance contributions are from effective local operators $Q_7$, $Q_9$ and $Q_{10}$, which come
 from the penguin and box diagrams.

In quantum chromodynamics (QCD), quarks are confined in color singlets. The $b\rightarrow s$ transition happens
inside hadrons. Therefore the matrix elements of the above three local operators have to be computed using 
non-perturbative methods, for example, lattice QCD. Those matrix elements can be parametrized by
form factors according to their Lorentz structures.
In total, there are ten form factors for the quark currents in $Q_7$, $Q_9$,
and $Q_{10}$, and our aim is to calculate these on the lattice with
dynamical simulations.

More details of our calculation strategy and our definitions of the form factors can be found in Ref.~\cite{Liu:2009dj}.
Here we update our progress in the extraction of the form factors. In the next section, we present our
lattice setup and then we show some preliminary results in the last section.

\section{Lattice setup}

We use configurations from the MIMD\footnote{Multiple Instruction stream, Multiple Data stream.} 
Lattice Computation (MILC) Collaboration,
which are $2+1$ flavour dynamical simulations 
using $\mathcal{O}(a^2)$ and tadpole improved staggered fermions (AsqTad)~\cite{Bazavov:2009bb}.
\begin{table}[hb]
\begin{center}
\begin{tabular}{cccccc}
\hline
\hline
       & $a$(fm) & $am_{sea}$ & Volume         & $N_{conf}\times N_{src}$ & $am_{val}$ \\
\hline
coarse & $\sim$0.12         & $0.007/0.05$  & $20^3\times64$ & $2109\times8$       & $0.007/0.04$ \\
       &              & $0.02/0.05$ & $20^3\times64$ & $2052\times8$ & $0.02/0.04$ \\
\hline
fine   & $\sim$0.09         & $0.0062/0.031$ & $28^3\times96$ & $1910\times8$ & $0.0062/0.031$ \\
\hline
\hline
\end{tabular}
\caption{Parameters of lattices being used in this study. $N_{src}$ is the number of point sources used on each
configuration.}
\label{tab:lat_parameters}
\end{center}
\end{table}
We currently have data from two
lattice spacings. At the coarse lattice spacing we have two different
light quark masses. The lightest
quark mass gives a pion mass of about 300 MeV. The parameters of our calculation are collected in 
Table~\ref{tab:lat_parameters}.
On each configuration, eight point sources are used to increase statistics. In~\cite{Liu:2009dj}, 
we found $Z_2\times Z_2$
random wall sources were inefficient for vector mesons and heavy-light mesons in reducing statistical errors of
correlators (random wall source methods allow one to approximately obtain all to all correlators and thus possibly to 
reduce statistical errors). Therefore we now use several point sources.

The light valance quarks are also AsqTad fermions as the sea quarks. For the heavy $b$ quark, we
use the (moving-)non-relativistic QCD (NRQCD) action~\cite{Horgan:2009ti}, which is expanded up to and including 
$\mathcal{O}(\Lambda^2_{QCD}/m_b^2)$. We can work directly at the physical $b$ quark
mass, thus no extrapolation up to $m_b$ is required.
In the lattice heavy-light currents, the heavy quark expansion includes order $1/m_b$.

We compute 2-point functions for the heavy-light $B$ meson and the light-light final state mesons as well
as 3-point functions with the current operators inserted. Then we fit these correlation functions with the
Bayesian fitting method~\cite{Lepage:2001ym}. From the fitted ground state energies and amplitudes, 
one can extract the
matrix elements of the current operators and then the form factors. The detailed formulas
can be found in Refs.~\cite{Liu:2009dj,Meinel:2008th}.

Matching factors of the lattice current operators to the $\overline{\rm MS}$ scheme were computed perturbative in 
Ref.~\cite{Muller:2010kb} to one loop. We set $\alpha_s=0.3$ below
to get the values of these matching factors.

Previous lattice calculations of the above form factors are all quenched calculations. And an extrapolation up
to the physical $b$ quark mass is needed for the heavy quark.
See, e.g.,~\cite{Becirevic:2006nm} and the references therein.

\section{Preliminary results}
\begin{figure}[p]
\centering
\includegraphics[height=2in,width=0.48\linewidth]{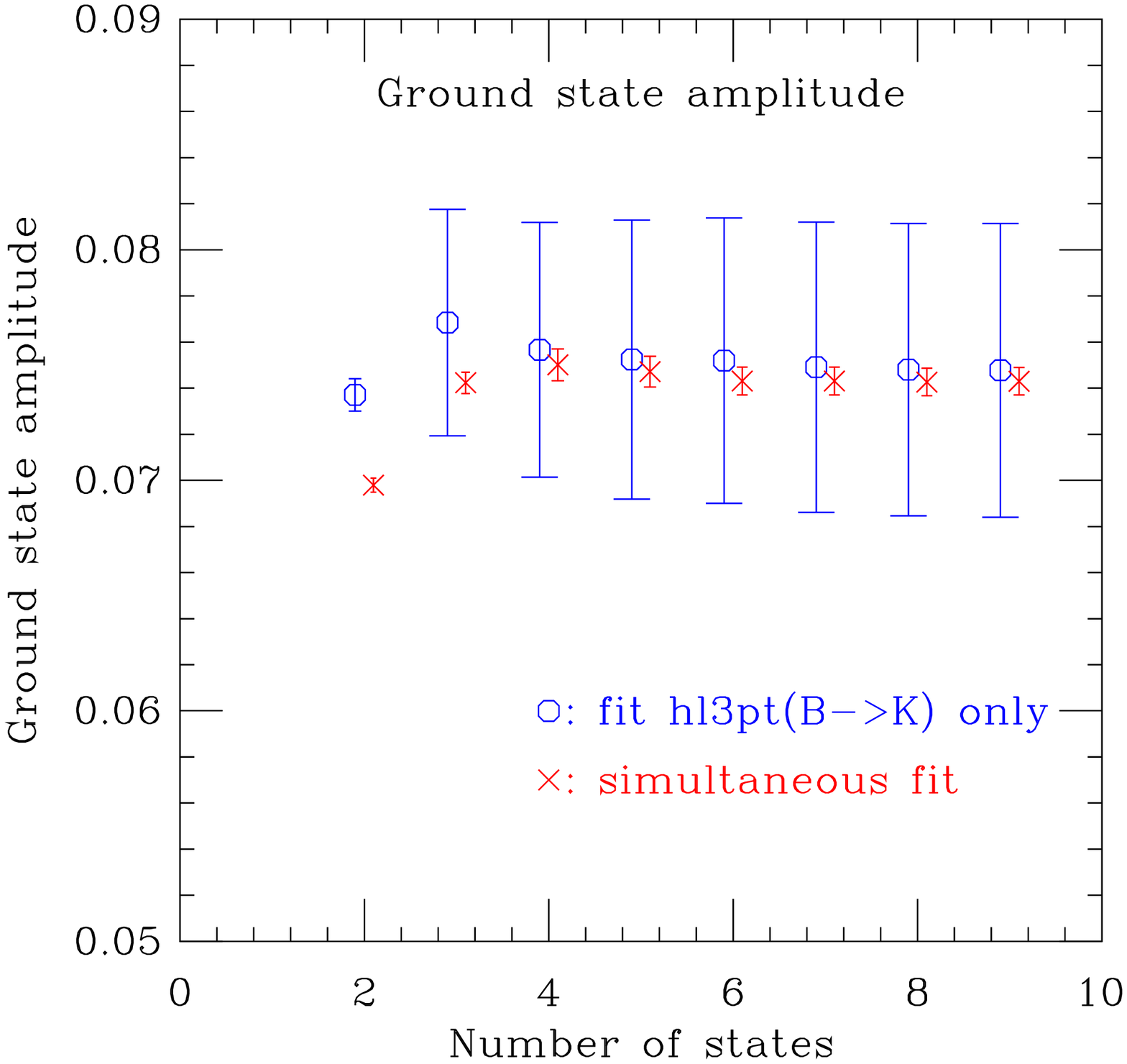}
\includegraphics[height=2in,width=0.48\linewidth]{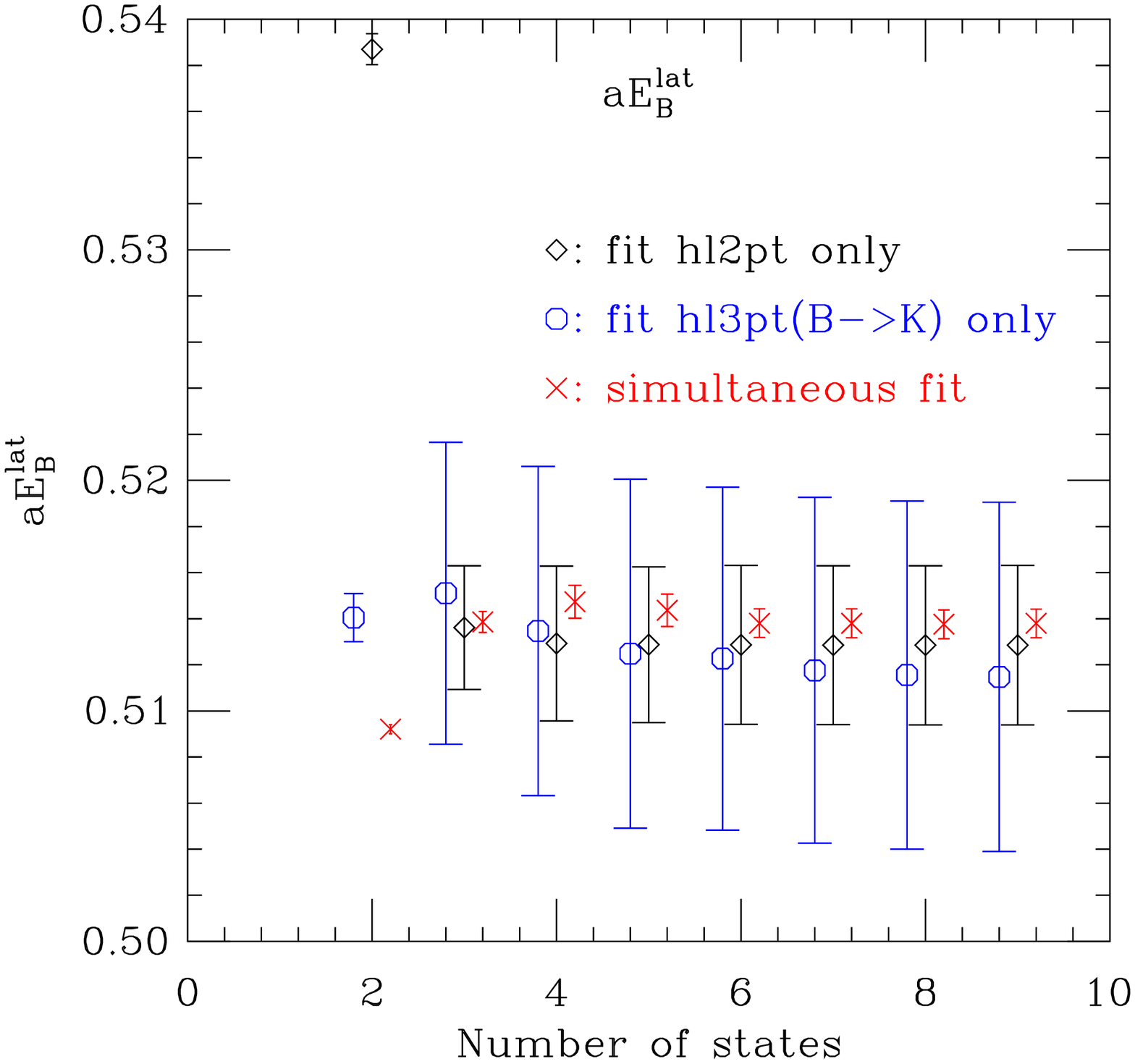}
\caption{Comparison of separate Bayesian fitting and simultaneous Bayesian fitting to heavy-light 2-point (hl2pt)
and heavy-light 3-point (hl3pt) functions
for the ground state amplitude in a 3-point function (left graph) and the $B$ meson energy $aE_B^{lat}$
(right graph) in lattice units. 
The horizontal axis is the number of (ground and excited) states in the fitting functions.
}
\label{fig:Aee_MB}
\end{figure}
We tried both separate fitting and simultaneous fitting of the 2- and 3-point functions.
The simultaneous fitting, where common
energy parameters are used, gives better results with smaller
statistical errors.
Examples are shown in Fig.~\ref{fig:Aee_MB} for the ground state amplitude in
a 3-point function for $B\rightarrow K$ and the $B$ meson
energy (in lattice units) obtained from different fits. Note a mass shifting term needs to be added for the
$B$ meson energy due to the use of (moving-)NRQCD.
In Fig.~\ref{fig:Aee_MB} the horizontal axis
is the number of (ground and excited) states in the Bayesian fitting functions. 
Equal number of opposite parity states are also in the fitting functions due to the use of
staggered fermions. The fitting results stabilize when more than 6
normal states and 6 opposite parity states are used.

\begin{figure}
\begin{minipage}[t]{.48\linewidth}
\centerline{\includegraphics[width=\linewidth,height=2in]{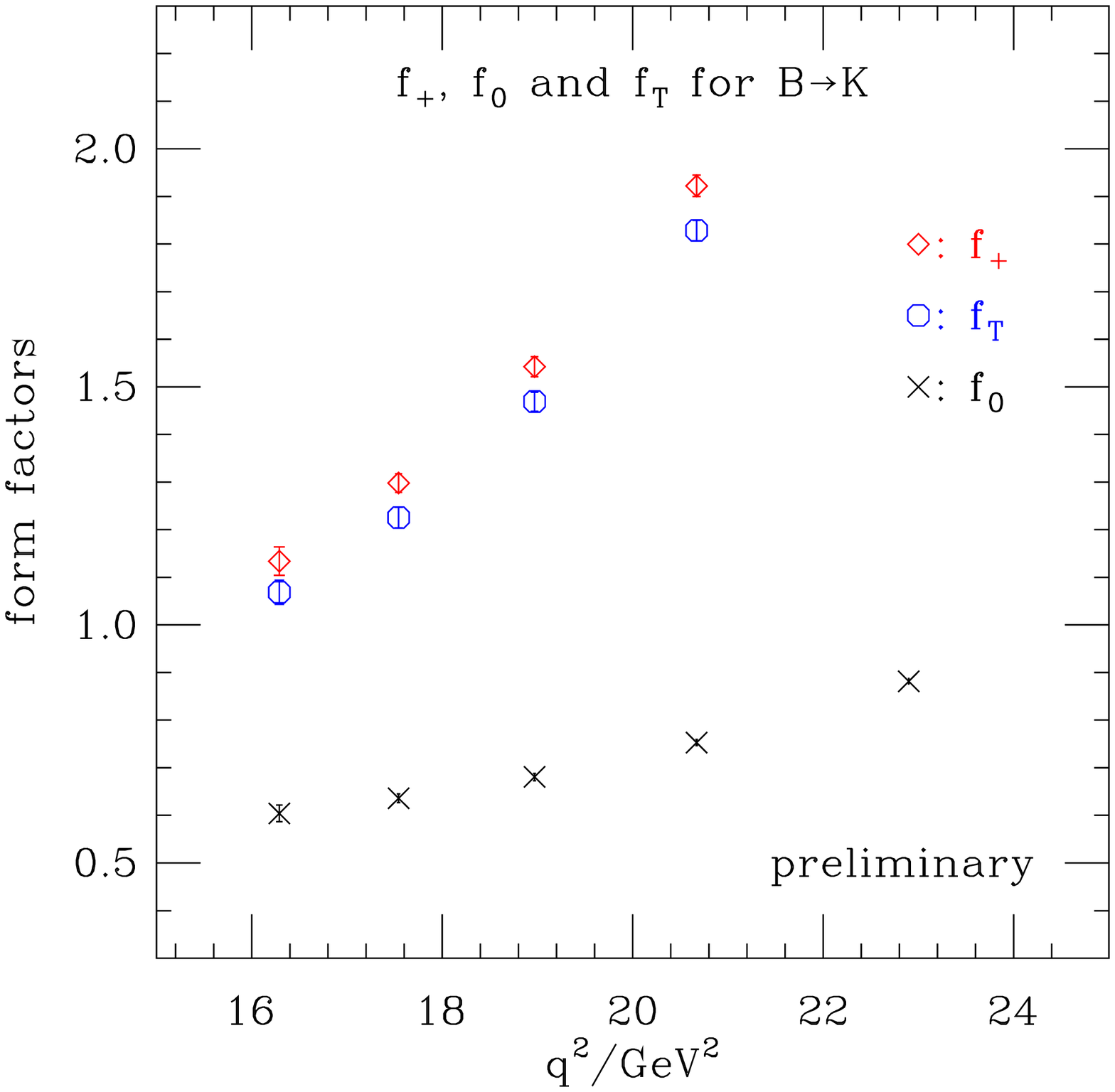}}
\caption{Preliminary results for the form factors $f_0$, $f_+$ and $f_T$ 
for $B\rightarrow K$ decays, obtained from simultaneous
Bayesian fits. The left-most points have $\vec v=(0,0,0)$,
$\vec p_B=0$ and $\vec p'_K=2\pi/L\cdot(-2,0,0)$.}
\label{fig:B_to_K_f0_fplus_fT}
\end{minipage}
\hfill
\begin{minipage}[t]{.48\linewidth}
\centerline{\includegraphics[width=\linewidth,height=2in]{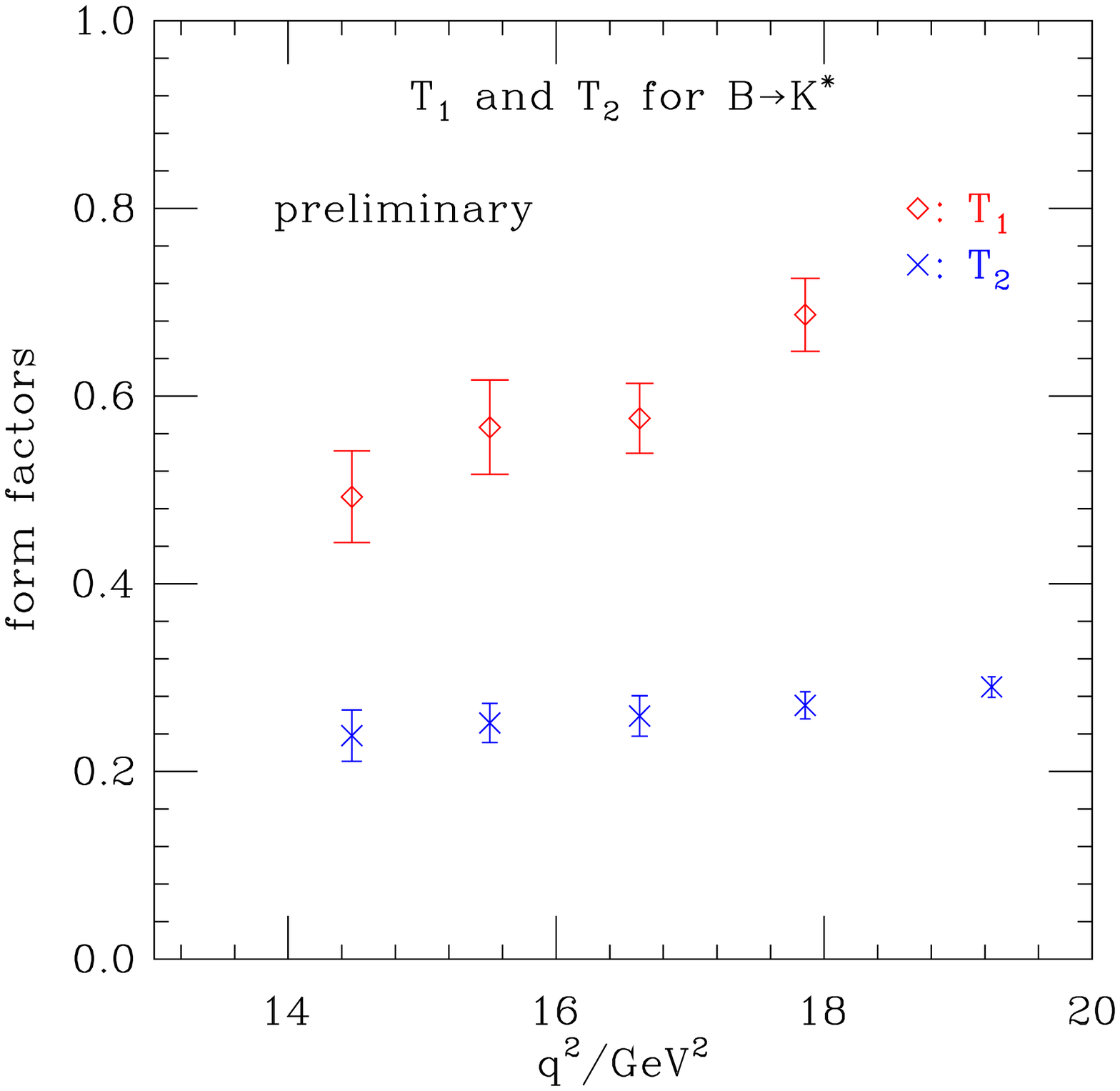}}
\caption{Preliminary results for the form factor $T_1$ and $T_2$ for $B\rightarrow K^*$ decays, 
obtained from simultaneous Bayesian fits. The left-most points have $\vec v=(0,0,0)$, 
$\vec p_B=0$ and $\vec p'_{K^*}=2\pi/L\cdot(-2,0,0)$.}
\label{fig:B_to_Kstar}
\end{minipage}
\end{figure}
Preliminary results of some form factors for $B\rightarrow K$ and $B\rightarrow K^*$ are given
in Fig.~\ref{fig:B_to_K_f0_fplus_fT} and Fig.~\ref{fig:B_to_Kstar} respectively. They are from the coarse
lattice with light valance quark masses $0.007/0.04$. The $1/m_b$ corrections to the currents are not included
yet, but will be soon. A first fitting to the 3-point functions shows that these corrections are small.
For the other quark mass and for the fine lattice spacing, 
data analysis is going on.
We will have more data points at lower $q^2$ after we analyze the correlators with non-zero velocity $\vec v$ for the
$B$ meson in the moving NRQCD action.

\begin{figure}[htb]
\centering
\includegraphics[height=2in,width=0.48\linewidth]{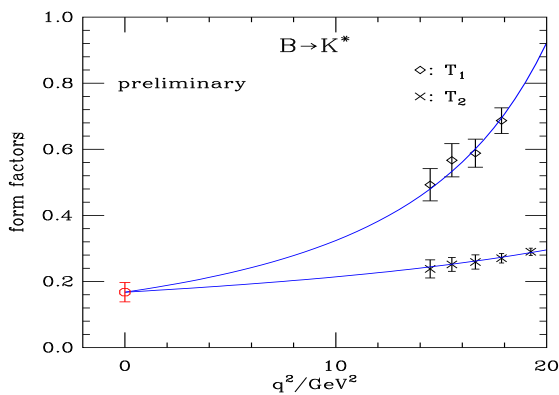}
\caption{Extrapolation of $T_1$ and $T_2$ to $q^2=0$ using the B-K ansatz in Eq.(\ref{eq:BK_ansatz}).}
\label{fig:T_zero}
\end{figure}

For the radiative decay $B\rightarrow \gamma K^*$, we want the form factor $T(0)=T_1(0)=T_2(0)$ to get
the branching fraction. To extrapolate our results to the $q^2=0$ limit, we need some ansatz. In Fig.~\ref{fig:T_zero} we 
do an extrapolation using the B-K ansatz~\cite{Becirevic:1999kt}
\begin{equation}
T_1(q^2)=\frac{T(0)}{(1-\tilde q^2)(1-\alpha\tilde q^2)},\quad
T_2(q^2)=\frac{T(0)}{1-\tilde q^2/\beta},\quad \tilde q^2=q^2/M^2_{B^*_s},
\label{eq:BK_ansatz}
\end{equation}
with $M_{B^*_s}=5.4158$ GeV fixed. 
This is only preliminary and we get $T(0)=0.168(29)$. The statistical errors will be reduced when 
more data points at low $q^2$ are included.

Eventually we need to extrapolate to the physical pion
mass and the continuum limit to really compare with experiments.
In a recent paper by H.~Na et al.\cite{Na:2010uf} on $D\rightarrow K$ form factors, 
the extrapolation in $q^2$, light quark mass and lattice spacing
were done all together using a ``simultaneous modified $z$-expansion".
We may try this method in the future.


\end{document}